\documentclass[12pt]{iopart}
\usepackage{graphicx}

\usepackage{iopams}
\usepackage{graphicx}
\usepackage{dcolumn}
\usepackage{CJK}
\usepackage{bm}
\usepackage[dvipdfm, pdfstartview=FitH, CJKbookmarks=true, bookmarksnumbered=true, bookmarksopen=true, colorlinks=true, pdfborder=001, citecolor=blue, linkcolor=blue, linktocpage=true] {hyperref}
\begin{document}

\title[Symmetry-Protected Quantum Adiabatic Evolution]{Symmetry-Protected Quantum Adiabatic Evolution in Spontaneous Symmetry-Breaking Transitions}

\author{Min Zhuang $^{1,2}$, Jiahao Huang $^{1}$, Yongguan Ke $^{1}$, and Chaohong Lee$^{1,2}$}

\address{$^1$ Laboratory of Quantum Engineering and Quantum Metrology, School of Physics and Astronomy, Sun Yat-Sen University (Zhuhai Campus), Zhuhai 519082, China}
\address{$^2$ Key Laboratory of Optoelectronic Materials and Technologies, Sun Yat-Sen University (Guangzhou Campus), Guangzhou 510275, China}

\ead{lichaoh2@mail.sysu.edu.cn}
\vspace{10pt}

\begin{abstract}
Quantum adiabatic evolution, an important fundamental concept in physics, describes the dynamical evolution arbitrarily close to the instantaneous eigenstate of a slowly driven Hamiltonian.
In most systems undergoing spontaneous symmetry-breaking transitions, their two lowest eigenstates change from non-degenerate to degenerate.
Therefore, due to the corresponding energy-gap vanishes, the conventional adiabatic condition becomes invalid.
Here we explore the existence of quantum adiabatic evolutions in spontaneous symmetry-breaking transitions and derive a symmetry-dependent adiabatic condition.
Because the driven Hamiltonian conserves the symmetry in the whole process, the transition between different instantaneous eigenstates with different symmetries is forbidden.
Therefore, even if the minimum energy-gap vanishes, symmetry-protected quantum adiabatic evolution can still appear when the driven system varies according to the symmetry-dependent adiabatic condition.
This study not only advances our understandings of quantum adiabatic evolution and spontaneous symmetry-breaking transitions, but also provides extensive applications ranging from quantum state engineering, topological Thouless pumping to quantum computing.
\end{abstract}

%
\vspace{2pc}

\section{Introduction}

Quantum adiabatic theorem (QAT) states that a slowly driven system from an initial eigenstate will stay close to the correspondingly instantaneous eigenstate of its Hamiltonian $H(t)$~\cite{PEhrenfest1916,MBorn1928,JSchwinger1937,TKato1950}.
The QAT is the theoretical basis for the Landau-Zener tunneling~\cite{Landau1932,Zener1932}, the perturbative quantum field theory~\cite{Gell-Mann1951}, the Berry phase~\cite{Berry1984},the topological Thouless pumping~\cite{Thouless1983} and the quantum annealing~\cite{Finnila1994,Johnson2011,Chancellor2017,Boixo2013} etc.
Moreover, the QAT has promising applications in quantum technologies such as quantum state engineering~\cite{Kral2007,Vitanov2017} and quantum computing~\cite{Albash2018}.
Usually, given the instantaneous eigenvalues $\{E_n(t)\}$ and eigenstates $\{|E_n(t)\rangle\}$ of $H(t)$, the QAT requires that $H(t)$ slowly varies according to $\left|i\hbar \langle E_n|\dot{E}_m\rangle\right| \ll \left| E_n- E_m\right|$ for $n \ne m$~\cite{KPMarzlin2004, DMTong2005, DMTong2007, GRigolin2010}.
If the energy degeneracy does not change, that is, the energy gap between neighboring energy eigenstates~\cite{KPMarzlin2004, DMTong2005, DMTong2007} (or neighboring degenerate energy eigenspaces~\cite{GRigolin2010}) is always open, this condition can always be satisfied if the driving is sufficiently slow.
However, it is still unclear whether there is an adiabatic condition for the slowly driven system involving degeneracy change.

Spontaneous symmetry-breaking (SSB) is a powerful fundamental concept in understanding continuous phase transitions~\cite{Sachdev2011}.
An SSB takes place when the ground state does not display a symmetry of the physical system.
In most systems undergoing SSB transitions, such as the transverse-field quantum Ising model~\cite{Sachdev2011, Zurek2005}, the Lipkin-Meshkov-Glick model~\cite{Lipkin1965,Yoshimura2014} and the quantized Bose-Josephson junction~\cite{Lee2006,Lee2009}, the two lowest eigenstates vary from non-degenerate to degenerate.
Therefore, driving a system through an SSB transition, the conventional adiabatic condition becomes invalid due to the corresponding energy gap vanishes.
Does this mean that the time-evolution dynamics is always non-adiabatic in such a driving process?
If quantum adiabatic evolution can still appear, what is the adiabatic condition?

In this article, we study the slow dynamics in a system driven through an SSB transition.
In the driving process, although the instantaneous ground states undergo an SSB, the driven Hamiltonian itself keeps the symmetry unchanged.
Thus, the population can only transfer between the instantaneous eigenstates of the same symmetry.
Based on this, we derive a symmetry-dependent adiabatic condition, which ensure the existence of symmetry-protected quantum adiabatic evolution even if the minimum energy-gap vanishes.
To illustrate our generic statements, we consider two typical examples: (i) the single-particle system within a symmetric one-dimensional potential varying from single-well to double-well, and (ii) the transverse-field quantum Ising model undergoing an SSB transition.

\section{Symmetry-protected transition and symmetry-dependent adiabatic evolution}
We consider a driven quantum system $\hat{H}(\textbf{R}(t))$ with a time-independent symmetry $\hat{Y}$ obeying the commutation relation $\left[\hat{H}(\textbf{R}(t)),\hat{Y}\right]=0$.
Generally, the Hamiltonian can be given as $\hat{H}(\textbf{R}(t))=\sum_{i=1}^{K} R_i(t)\hat{H}_i$ with the time-varying parameters $\textbf{R}(t)=[R_1 (t),R_2 (t),\cdots,R_K(t)]$ and the time-independent operators $\hat{H}_i$.
%
%
Thus, an arbitrary state can be expanded by the instantaneous simultaneous eigenstates of $\hat{Y}$ and $\hat{H}(\textbf{R}(t))$: $\{|{\phi_{E_n}^{\lambda_{\alpha}} (\textbf{R}(t))}\rangle\}$.
Here, $E_{n}$ and $\lambda_{\alpha}$ stand for the $n$-th eigenvalue of $\hat{H}(\textbf{R}(t))$ and the $\alpha$-th eigenvalue of $\hat{Y}$, respectively.

\subsection{Symmetry-protected transition} 

As the symmetry $\hat{Y}$ is a time-independent operator, we have $\frac{\partial{}}{\partial{t}}\left[\hat{Y} {|{\phi_{E_n}^{\lambda_{\alpha}} (\textbf{R}(t))}}\rangle\right]=\hat{Y}\frac{\partial{}}{\partial{t}} {|{\phi_{E_n}^{\lambda_{\alpha}} (\textbf{R}(t))}\rangle}$ and their inner products (with ${\langle{\phi_{E_m}^{\lambda_{\beta}} (\textbf{R}(t))}|}$) satisfying
%
\begin{eqnarray}\label{Eq:partial_Invarient}
{\langle{\phi_{E_m}^{\lambda_{\beta}} (\textbf{R}(t))}}|{\frac{\partial}{\partial{t}}}|
 {{{\phi}_{E_n}^{\lambda_{\alpha}} (\textbf{R}(t))}}\rangle(\lambda_{\beta}-\lambda_{\alpha})=0.
\end{eqnarray}
Due to $i\hbar \frac{\partial}{\partial{t}} =\hat{H}(\textbf{R}(t))$, we have
\begin{eqnarray}\label{Eq:SPT}
H_{mn}^{\beta \alpha}(t)(\lambda_{\beta}-\lambda_{\alpha})=0
\end{eqnarray}
with $H_{mn}^{\beta \alpha}(t) = {\langle{\phi_{E_m}^{\lambda_{\beta}}(\textbf{R}(t))}} |\hat{H}(\textbf{R}(t))| {{\phi_{E_n}^{\lambda_{\alpha}}(\textbf{R}(t))}\rangle}$(see appendix \emph{A}).
This means that the state transition is protected by the symmetry.
For the instantaneous eigenstates of the same symmetry (i.e. ${\lambda}_{\beta} = {\lambda}_{\alpha}$), the population may transfer between them.
For the instantaneous eigenstates with different symmetries (i.e. ${\lambda}_{\beta} \neq {\lambda}_{\alpha}$), even if their instantaneous eigenenergies are degenerate, the population transfer between them is exactly forbidden.
The symmetry-protected transition have been mentioned in discussing the dynamics crossing through quantum phase transitions~\cite{Yoshimura2014,Dziarmaga2005,Quan2010,Dziarmaga2010}.
In the following, we make use of this property and explore the symmetry-dependent adiabatic evolution.

\subsection{Symmetry-dependent adiabatic evolution} 

The symmetry-protected transition is the basis for the following symmetry-dependent adiabatic condition (SDAC).
Without loss of generality, we derive the SDAC for degenerate systems, which can be relaxed to the non-degenerate systems.
Below, $\mathbb{H}_{\emph{m}}(\textbf{R}(t))$ denotes the $m$-th degenerate subspace of $\hat{H}(\textbf{R}(t))$ with the eigenenergy $E_m$ and the degeneracy number $d_m$.

We assume the system is driven from an instantaneous eigenstate ${|{\phi_{E_m}^{\lambda_{\beta}} (\textbf{R}(t))}\rangle}$ in a degenerate subspace $\mathbb{H}_{\emph{m}}(\textbf{R}(t))$, in which each eigenstate has different symmetry (i.e. $\lambda_{i}\neq\lambda_{j}$ if $i\neq j$ for $i,j=\{1,2,\cdots,d_m\}$).
Thus, the adiabatic condition for remaining in the same instantaneous eigenstate at time $t+dt$ (where $dt$ is an infinitesimal interval) is given as
\begin{eqnarray}\label{Eq:AC}
\epsilon(t)=\max_{\{n\}}\left\{ \left| \frac{H_{mn}^{\beta\beta}(t)}{E_m-E_n} \right|\right\} \ll 1~~\textrm{with}~~m\neq n,
\end{eqnarray}
where $H_{mn}^{\beta\beta} (t) = i\hbar {\langle{\phi_{E_{m}}^{\lambda_{\beta}} (\textbf{R}(t))}|
{\dot{\phi}_{E_{n}}^{\lambda_{\beta}} (\textbf{R}(t))}\rangle}$, $\lambda_{\beta}$ denotes the symmetry and $E_{\{m,n\}}$ stand for the instantaneous eigenenergies (see the detailed derivation in the appendix \emph{B}).
Since $E_m$ and $E_n$ belong to different subspaces $\mathbb{H}_{\emph{m}}(\textbf{R}(t))$ and $\mathbb{H}_{\emph{n}}(\textbf{R}(t))$, the energy gap does not vanish, i.e., $|E_m-E_n| > 0$.
This condition implies that, the adiabaticity of the time-evolution is determined by the energy gap between neighboring instantaneous eigenstates of the same symmetry.
Thus there is no transition between eigenstates with different symmetries even if their energy gap vanishes.
When $d_m=1$, the subspace $\mathbb{H}_{\emph{m}}(\textbf{R}(t))$ becomes non-degenerate, and the above SDAC keeps valid.
If there is no symmetry-dependent behavior, that is, all $\lambda_{\alpha}$ have the same value, the SDAC becomes the conventional adiabatic condition.
According to the SDAC~(\ref{Eq:AC}), adiabatic evolutions may still appear even if the energy gap between nearest neighboring eigenstates vanishes.
In a driven system through an SSB transition, in which the two lowest eigenstates change from non-degenerate to degenerate, the dynamics may still evolve arbitrarily close to its instantaneous ground state if there is a finite minimum energy gap between instantaneous eigenstates of the same symmetry.
Naively, one can drive a system parameter $R$ linearly with fixed sweeping rate $\dot{R}=\upsilon$ from the non-degenerate regime across to the degenerate regime. If $\upsilon$ is sufficiently small, the adiabatic evolution of the ground state can still be achieved with high fidelity.
However, this linear sweeping scheme is not timesaving.

To perform fast and efficient ground state adiabatic evolution, we propose to change the sweeping rate with time according to the instantaneous energy gaps between the eigenstates of the same symmetry [designed by SDAC~(\ref{Eq:AC})], i.e.,
\begin{eqnarray}\label{Eq:Sweeping_rate}
\upsilon(t)\!=\!\min_{\{n\}}\!\left\{\!\frac{{\epsilon\left[E_{1}(R(t))-E_{n}(R(t))\right]^{2}}}{\hbar\left|\langle{\phi_{E_{1}}^{\lambda_{\beta}}(R(t))}|\!
{\frac{\partial{\hat{{H}}(R(t))}}{\partial {R(t)}}}\!|
{{\phi_{E_{n}}^{\lambda_{\beta}}(R(t))}}\rangle\right|}\right\}
\end{eqnarray}
with the fixed adiabatic parameter $\epsilon$.
Thus, this scheme is called $\epsilon$-fixed sweeping.
We define the fidelity $F_{n}^{\alpha}(t)=|\langle\Psi(x,t)|\phi_{E_{n}}^{\lambda_{\alpha}}\rangle|^2$ between the instantaneous evolved state $|{\Psi(x,t)}\rangle$ and the $n$-th instantaneous eigenstate $|{\phi_{E_{n}}^{\lambda_{\alpha}}}\rangle$.
If the system evolves from its instantaneous ground state $|\phi_{E_{1}}^{\lambda_{\beta}}\rangle$, smaller $\epsilon$ corresponds to higher fidelity $F_{1}^{\beta}(t)$.
In the limit of small $\epsilon$ (i.e. $\epsilon\ll1$), we analytically obtain an inequality between the fidelity of staying in the instantaneous ground state and the adiabatic parameter (see appendix C),
\begin{equation}\label{Eq:a1}
F_{1}^{\beta}(t)\gtrsim\left(1\!-\!\frac{8{\epsilon}^2}{1+4{\epsilon}^2}\right)^2\approx 1-16\epsilon^2+O(\epsilon^4).
\end{equation}

\section{Single-particle system within a symmetric potential varying from single-well to double-well}

We first consider a single particle confined within a symmetric one-dimensional potential, which slowly varies from single-well to double-well.
Its Hamiltonian reads,
\begin{equation}\label{Eq:Ha_Single}
\hat{H}_{\textrm{S}}(x,t) = -\frac{{\hbar}^{2}}{2m}\frac{{\partial}^2}{{\partial{x}^2}} +V(x,t).
\end{equation}
%
%
The first term is the kinetic energy and the second term describes the potential.
The time-varying potential $V(x,t)$ is a superposition of a time-independent harmonic trap and a time-dependent Gaussian barrier, $V(x,t)=\frac{1}{2}m\omega^2 {x}^{2}+A(t)e^{-\frac{{x}^2}{2d^{2}}}$.
%
%
Here, $\omega$ is the trapping frequency, $d$ denotes the barrier width and the barrier height $A(t)$ varies with time.
At $t=0$, $A=0$, $V(x,t)$ is a harmonic potential (a symmetric single-well potential).
When $A(t)$ increases with time, $V(x,t)$ gradually becomes a symmetric double-well potential, and the two lowest eigenstates change from non-degenerate to degenerate (or quasi-degenerate for a large but finite $A(t)$).

\begin{figure}
  \centering\includegraphics[width=0.6\columnwidth]{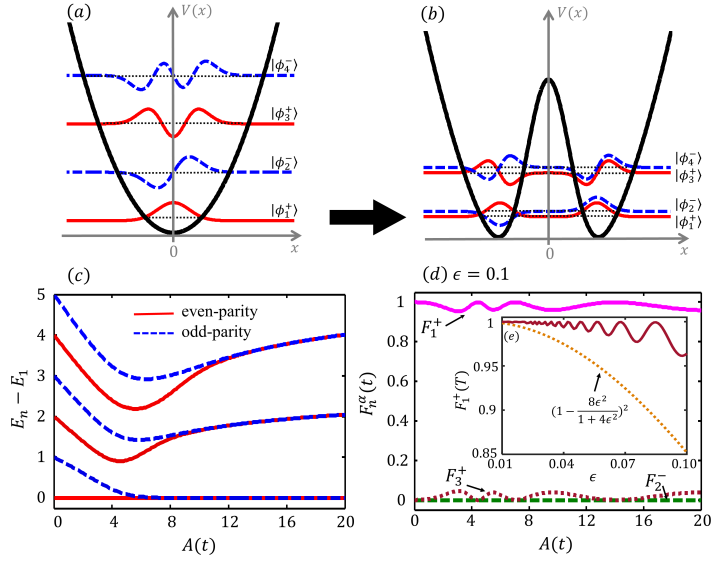}
  \caption{\label{Fig1} The single-particle system within a symmetric one-dimensional potential varying from (a) single-well to (b) double-well. The red-solid and blue-dashed lines stand for even and odd parities, respectively. (c) The energy spectrum versus the barrier height $A$. (d) The populations in the lowest three instantaneous eigenstates $F_1^{+}$, $F_2^{-}$, $F_3^{+}$ versus the barrier height $A(t)$ for $\epsilon=0.1$. (e) The final population in the instantaneous ground state $F_1^{+}(T)$ (where $A=20$) versus $\epsilon$.
  }
\end{figure}

In the whole process, the Hamiltonian~(\ref{Eq:Ha_Single}) keeps the mirror-reflection parity symmetry.
That is, $\hat{H}_{\textrm{S}}(x,t)$ is invariant under the mirror reflection $\hat{P}: \!x\rightarrow \!\!-x$, $\left[\hat{H}_{\textrm{S}}(x,t),\hat{P}\right]=0$.
Thus $\hat{P}$ have two eigenvalues $\pm1$ respectively representing even and odd parity.
Due to the parity symmetry, the instantaneous eigenstates appear with even and odd parity alternately.
Initially, the energy levels are non-degenerate, see Fig.~\ref{Fig1}~(a).
When the barrier height is sufficiently high (i.e. $A\gg {\omega}^2 {d}^2$), the neighbouring pairs of eigenstates of different parity become quasi-degenerate, see Fig.~\ref{Fig1}~(b).
The quasi-degeneracy is also evidenced by the static energy spectrum versus the barrier height $A$, see Fig.~\ref{Fig1}~(c).

Now we discuss how adiabatic evolution appears.
Due to the symmetry protected transition, from an initial even-parity eigenstate, the odd-parity instantaneous eigenstates will never be populated and vice versa.
According to the SQAC~(\ref{Eq:AC}), the adiabaticity is determined by the minimum energy gap between the instantaneous eigenstates of the same symmetry.
Since there always exists a finite gap between the instantaneous eigenstates of the same symmetry, adiabatic evolution may always appear if the system is driven sufficiently slowly.

To show how to achieve adiabatic evolution via designing the sweeping process of the barrier height, we perform a numerical calculation based upon the Schr\"{o}dinger equation $i\hbar \frac{\partial}{\partial t} |{{\Psi}(x,t)}\rangle = \hat{H}_{\emph{S}}(x,t)|{\Psi(x,t)}\rangle$.
In our calculation, we use the dimensionless units of $m=\hbar=\omega=1$ and set $d=\sqrt{2}$.
We choose the initial state as the ground state $|{\phi_{E_1}^+}\rangle$ of $\hat{H}_S(0)$ with even parity.
The barrier height is gradually lifted from $A(0)=0$ to $A(T)\gg {\omega}^2 {d}^2$.
Since the nearest eigenstate with even parity is the second-excited state, the SDAC in this system is
\begin{equation}\label{SDAC-S}
\epsilon(t)=\left|{\frac{\hbar\langle{\phi_{E_{1}}^{+}(A(t))}|v_{A}(t)
{e^{{-{x}^{2}}/{2d}}}|{{\phi_{E_{3}}^{+}(A(t))}}\rangle}{\left[E_{1}(A(t))-E_{3}(A(t))\right]^{2}}}\right| \ll 1.
\end{equation}
Thus, the sweeping process is described as $A(t)=\int_{0}^{t}v_{A}(t')dt'$ with the sweeping rate $v_{A}(t)$.
Given $\epsilon$, according to the Eq.~(\ref{Eq:Sweeping_rate}), we have $v_{A}(t) = \frac{\epsilon{\left[E_{1}(t)-E_{3}(t)\right]}^{2}} {\left|\langle{\phi_{E_{1}}^{+}(t)}|{e^{{-{x}^{2}}/{2d}}}| {\phi_{E_{3}}^{+}(t)}\rangle\right|}$.

\begin{figure}[!htp]
   \centering
   \includegraphics[width=0.6\columnwidth]{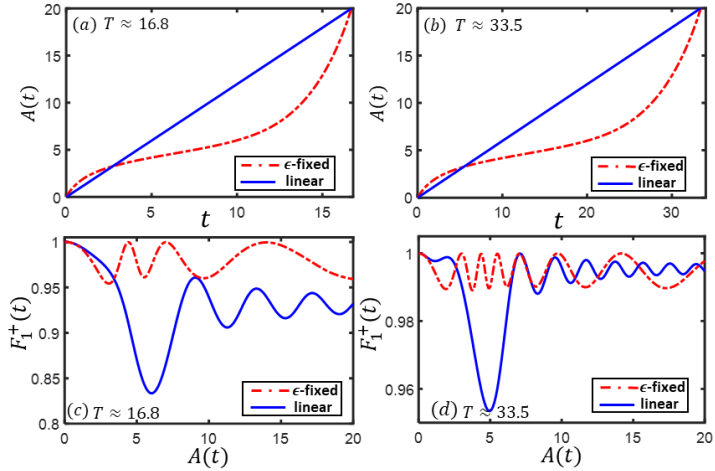}
  \caption{\label{Fig2} (a),(b) The changes of barrier height $A(t)$ with time for gap-dependent sweeping and linear sweeping. (c), (d) The comparison between our gap-dependent sweeping and the linear sweeping within the same evolution duration $T$. The evolved population in the instantaneous ground state ${F}_{1}$ when (c) $T\approx 17$ and (d) $T\approx 33$, respectively.
  }
\end{figure}

The time-evolution sensitively depends on the value of $\epsilon$.
In Fig.~\ref{Fig1}~(d), for $\epsilon=0.1$, we show the fidelities versus the instantaneous barrier height $A(t)$.
Although the first gap $E_2(t)-E_1(t)$ gradually vanishes, because the transition is protected by the symmetry, the population in the first-excited state $F_{2}^{-}$ keeps zero during the whole process.
Particularly, the population in the ground state $F_{1}^{+}$ keeps above $0.95$ and only small population is transferred to the second-excited state (characterized by $F_{3}^{+}$).
To show how slow the sweeping is practical, we plot the final fidelity $F_1^{+}(T)$ [where $A(T)=20$] versus the adiabatic parameter $\epsilon$, see Fig.~\ref{Fig1}~(e).
The final fidelity shows the appearance of adiabatic evolution for sufficiently small $\epsilon$.
Clearly, the curve $F_1^{+}(T)$ is always above the analytical line $(1\!-\!\frac{8{\epsilon}^2}{1+4{\epsilon}^2})^2$, which confirms the validity of (\ref{Eq:a1}).
%
%

Meanwhile, we compare the linear sweeping scheme and our $\epsilon$-fixed sweeping scheme under the same evolution duration $T$.
For the linear sweeping, $A(t)=v_A t$, as shown in Fig.~\ref{Fig2}~(a) and (b) (blue solid lines).
While for $\epsilon$-fixed sweeping, $v_{A}(t) = \frac{\epsilon{\left[{E_{1}}(t)-E_{3}(t)\right]}^{2}} {\left|\langle{\phi_{E_{1}}^{+}(t)}|{e^{{-{x}^{2}}/{2d}}}| {\phi_{E_{3}}^{+}(t)}\rangle\right|}$, as shown in Fig.~\ref{Fig2}~(a) and (b) (red dashed lines).
Under the same total duration $T$, the $\epsilon$-fixed sweeping outperforms the linear sweeping with larger final fidelity of staying in the ground state, see in Fig.~\ref{Fig2}~(c) and (d).
Besides, the amplitude of the oscillation of $F_1^{+}(t)$ with $\epsilon$-fixed sweeping is much smaller than linear sweeping.

\begin{figure}[!htp]
 \includegraphics[width=0.6\columnwidth]{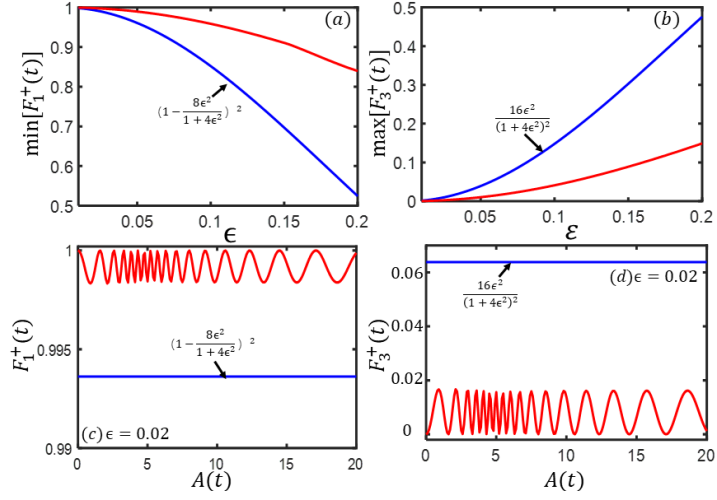}
   \centering\caption{\label{Fig3} The dynamical evolution of the single-particle system within a symmetric one-dimensional potential varying from single-well
to double-well. (a) The minimal fidelity in the instantaneous ground state $\min[F_1^{+}(t)]$ (gap-dependent sweeping) versus the adiabatic parameter $\epsilon$.
  The blue solid lines denote the analytic results for $\min[F_1^{+}(t)]$; (b) The maximum fidelity in the instantaneous second excited state $\max[F_3^{+}(t)]$ (gap-dependent sweeping) versus the adiabatic parameter $\epsilon$.
  The blue solid lines denote the analytic results for $\max[F_3^{+}(t)]$.
  (c) The evolution of $F_1^{+}(t)$ for $\epsilon=0.02$. (d) The evolution of $F_3^{+}(t)$ for $\epsilon=0.02$.
  }
\end{figure}

In Fig.~\ref{Fig3}~(a) and (b), we show the minimal fidelity $\min[F_{1}^{+}(t)]$ and the maximum fidelity $\max[F_{3}^{+}(t)]$ versus $\epsilon$.
In Fig.~\ref{Fig3}~(c) and (d), we give the time-evolution of $F_{1}^{+}(t)$ and $F_{3}^{+}(t)$ in the case of $\epsilon=0.02$.
The fidelity $F_{1}^{+}(t)$ and the fidelity $F_{3}^{+}(t)$ satisfy the two inequations, i.e., $ F_{1}^{+}(t)\geq \left(1-\frac{8{\epsilon}^2}{1+4{\epsilon}^2}\right)^{2}$ and $F_{3}^{+}(t) \leq \frac{16{\epsilon}^2}{(1+4{\epsilon}^2)^2}$.

\section{Transverse-field quantum Ising model driven through an SSB transition}

In addition to single-particle systems, symmetry-protected quantum adiabatic evolutions may also appear in many-body quantum systems driven through an SSB transition.
Below we consider the quantum Ising model: $\hat{H}_{Ising} (t) = \frac{B}{2} \sum ^{N}_{i=1}\sigma ^{x}_{i}+\sum^{N-1}_{i<j}\frac{J_{ij}}{2}\sigma ^{z}_{i}\sigma ^{z}_{j}$
with the Pauli operators $\sigma^{x,z}_{i}$, the Ising interaction $J_{ij}$, the transverse magnetic field ${B(t)}$ and the total spin number $N$.

This model is invariant under the transformation: $\sigma^x_{i}\rightarrow{\sigma^x_{i}}, \sigma^y_{i}\rightarrow{-\sigma^y_{i}}, \sigma^z_{i}\rightarrow{-\sigma^z_{i}}$.
By defining the parity operator, $\hat{P}=e^{-i\pi/2\sum_{i}\sigma ^{x}_{i}}$ for even $N$ and $\hat{P}=-ie^{-i\pi/2\sum_{i}\sigma ^{x}_{i}}$ for odd $N$~\cite{Yoshimura2014,Choi2017}, which has two eigenvalues $\pm1$ respectively representing even and odd parity, we have $\left[\hat{H}_{Ising},\hat{P}\right]=0$.
If the Hamiltonian is dominated by the first term, the ground state is a paramagnetic state of all spins aligned along the magnetic field $B$.
If the Hamiltonian is dominated by the second term and $J_{ij} < 0$, there appear two degenerate ferromagnetic ground states of all spins in either spin-up or spin-down.
Thus any superposition of these two ground states is also a ground state, in which the equal-probability superposition of these two states is known as a GHZ state.

\begin{figure}
  \centering\includegraphics[width=0.6\columnwidth]{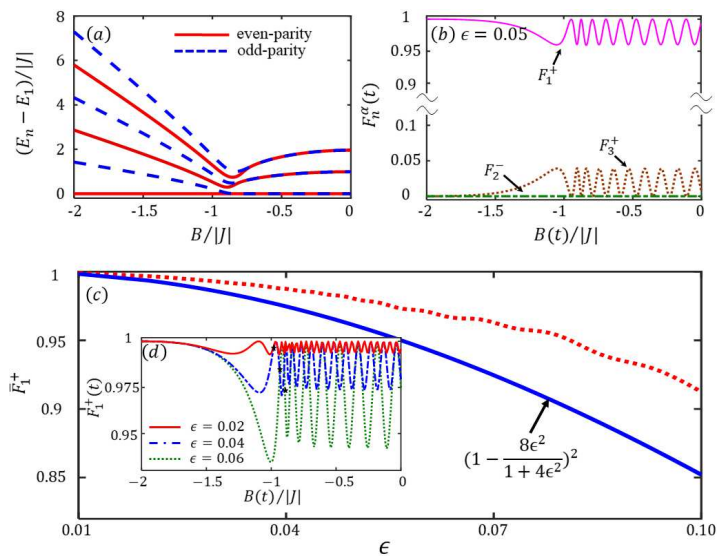}
  \centering\caption{\label{Fig4} The dynamics of the quantum Ising model~(\ref{Ising}) with $N=100$ driven through an SSB transition. (a) The energy spectrum. (b) The populations in the lowest three instantaneous eigenstates $F_1^{+}$, $F_2^{-}$, $F_3^{+}$ versus the magnetic field $B(t)$ for $\epsilon=0.05$. (c) The final average fidelity ${F}_1^{+}$ versus $\epsilon$, in which the star ($\star$) denotes the starting point of the average. The inset (d) shows the population dynamics in the instantaneous ground state $F_1^{+}$ for different $\epsilon$. The blue solid line denotes the analytic lower bound of $F_1^{+}(t)$.}
\end{figure}

The quantum Ising model has been experimentally realized via ultracold trapped ions~\cite{Porras2004}.
Generally, the Ising interaction is in form of $J_{ij}=J/|i-j|^{\delta}$ (with $0 \le \delta \le 3$)~\cite{Britton2012, Bohnet2016, Garttner2017}.
Recently, the homogeneous Ising interaction (i.e. $\delta=0$) and the time-varying transverse field $B(t)$ have been realized in experiments~\cite{Bohnet2016, Garttner2017}, that is, $J_{ij}=J/N$ and the Hamiltonian
\begin{equation}\label{Ising}
\hat{H}_{Ising}(t) = \frac{B(t)}{2} \sum^{N}_{i=1} \sigma^{x}_{i} + \frac{J}{2N} \sum^{N-1}_{i<j}\sigma ^{z}_{i}\sigma ^{z}_{j},
\end{equation}
which is equivalent to a symmetric Bose-Josephson junction~\cite{Lee2006,Lee2009}.
Below we concentrate on discussing the system of $J<0$, in which an SSB occurs at the critical point $|B_c/{J}|=1$ when $N \rightarrow \infty$.
Accompanying with the SSB, the two lowest eigenstates change from non-degenerate to degenerate.
Initially, the system is dominated by the transverse magnetic field (i.e. $|B|\gg |J|$), the energy levels are non-degenerate, and the eigenstates alternately appear with even and odd parity.
Fixing the Ising interaction, when the transverse magnetic field $B(t)< B_c$, the neighboring pairs of eigenstates $\{ {|{\phi_{E_{2n-1}}^{+}(t)}\rangle},{|{\phi_{E_{2n}}^{-}(t)}\rangle} \}$ ${(n=1,2,3....)}$ of different parity become degenerate (or quasi-degenerate for finite $N$).
It is worth to mention that, the minimum energy gap between the ground state and the second excited state is inversely proportional to the cube root of particle number, i.e. $E_3-E_1\propto N^{-1/3}$~\cite{Yoshimura2014, Botet1983}.
When $N \rightarrow \infty$, the gap $E_3-E_1 \rightarrow 0$ also vanishes, the non-adiabatic excitation will inevitably occur, which is consistent with the studies on Kibble-Zurek mechanism in quantum Ising model~\cite{Dziarmaga2005, Quan2010} and LMG model~\cite{Dziarmaga2010}.

\begin{figure}
  \centering\includegraphics[width=0.6\columnwidth]{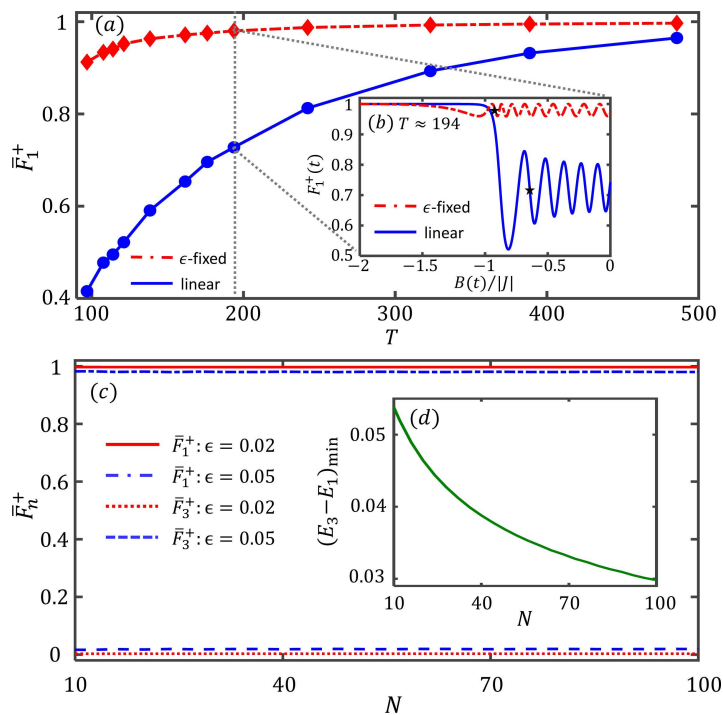}
  \centering\caption{\label{Fig5} The comparison between the $\epsilon$-fixed sweeping and the linear sweeping of the quantum Ising model~(\ref{Ising}). (a) The final average fidelity ${F}_1^{+}$ for $N=100$ versus the total evolution duration $T$. The inset (b) shows the time-evolution of $F_{1}^{+}$ with $T\approx194$, in which the star ($\star$) denotes the starting point of the average. (c) The final average fidelities versus $N$. (d) The minimum energy gap between the ground state and the second-excited state versus $N$.}
\end{figure}

For a realistic system, as its $N$ is finite, according to the SDAC~(\ref{Eq:AC}), 
adiabatic evolutions may still appear due to there always have finite energy gaps between eigenstates of the same parity.
In Fig.~\ref{Fig4} (a), we show the energy spectrum and the population dynamics for $N=100$.
The state is initialized as the ground state $|{\phi_{E_1}^+}\rangle$ of $\hat{H}_{T}(0)'$ with even parity.
Since the nearest eigenstate with even parity is the second-excited state, the SDAC in this system is
\begin{equation}\label{SDAC-S}
\epsilon(t)=\left|{\frac{\hbar\langle{\phi_{E_{1}}^{+}(A(t))}
|\frac{1}{2}v_{B}(t)\left(\sum ^{N}_{i=1} \sigma ^{x}_{i}\right)|{{\phi_{E_{3}}^{+}(A(t))}}\rangle}{\left[E_{1}(A(t))-E_{3}(A(t))\right]^{2}}}\right| \ll 1.
\end{equation}
Thus, the transverse magnetic field is gradually varied from $B(0)=-2$ to $B(T)=0$ across the transition point $B=1$.
The sweeping process is described as $B(t)=\int_{0}^{t}v_{B}(t')dt'$ with the sweeping rate $v_{B}(t) = \frac{\epsilon{\left[E_{1}(t)-{E_{3}}(t)\right]}^{2}} {\left|\langle{\phi_{E_{1}}^{+}(t)}|\frac{1}{2}\left(\sum ^{N}_{i=1} \sigma ^{x}_{i}\right) |{\phi_{E_{3}}^{+}(t)}\right|\rangle}$ determined according to Eq.~(\ref{Eq:Sweeping_rate}).

In Fig.~\ref{Fig4} (b), given $\epsilon=0.05$, we show the fidelities versus the instantaneous magnetic field $B(t)$.
Due to the two lowest instantaneous eigenstates have different symmetry and the system is driven from the ground state, as a result of the symmetry-protected transition, the instantaneous first-excited state is never occupied, see $F_{2}^{-}(t)$ in Fig.~\ref{Fig4}~(b).
Most of population stays in the instantaneous ground state and only small amount jumps to the instantaneous second-excited state, see $F_1^{+}(t)$ and $F_3^{+}(t)$ in Fig.~\ref{Fig4}~(b).
In general, the final fidelities (at B=0) oscillate with adiabatic parameter. To eliminate the oscillation and smooth the curve monotonously, we analyze the final average fidelity $\bar{F}_n^{\alpha}=\frac{1}{T-T^{*}}\int_{T^{*}}^T F_{n}^{\alpha}(t) dt$ with $T^{*}$ denoting the instant that the middle point of the first oscillation after the transition point.
This can be used as an indicator for finding out how slow the sweeping is practical.
In Fig.~\ref{Fig4}~(c), we show $\bar{F}_{1}^{+}$ versus $\epsilon$.
Clearly, $\bar{F}_{1}^{+}$ is always above the analytical lower bound (the solid blue line) given by~(\ref{Eq:a1}).

Here, we mention some advantages of our $\epsilon$-fixed sweeping scheme.
Firstly, the adiabaticity of our $\epsilon$-fixed sweeping is better than the one of the linear sweeping scheme.
For the same time-evolution duration $T$, our scheme always has a higher final average fidelity $\bar{F}_1^{+}$, see Fig.~\ref{Fig5}~(a) and (b).
Secondly, for a given adiabatic parameter $\epsilon$, the final average fidelities $\bar{F}_1^{+}$ and $\bar{F}_3^{+}$ remain almost the same for different $N$, see Fig.~\ref{Fig5}~(c).
Although the energy gap decreases with the total spin number $N$ [see Fig.~\ref{Fig5}~(d)], the final average fidelities are alomost independent on $N$ for a given $\epsilon$ [see Fig.~\ref{Fig5}~(d)].

\begin{figure}[!htp]
 \includegraphics[width=0.6\columnwidth]{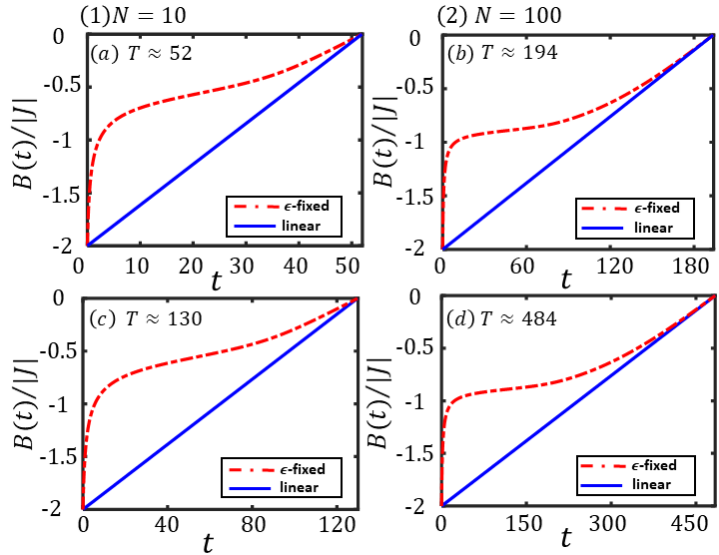}
   \centering\caption{\label{Fig6} The changes of the magnetic field amplitude $B(t)$ with time for $\epsilon$-fixed sweeping and linear sweeping. (a),(b) When $\epsilon=0.05$, the total evolution duration is $T\approx52$ for $N=10$ and $T\approx194$ for $N=100$. (c), (d) When $\epsilon=0.02$, the total evolution duration is $T\approx130$ for $N=10$ and $T\approx484$ for $N=100$.}
\end{figure}

In Fig.~\ref{Fig6}, we show the change of $B(t)$ for the linear sweeping (blue solid lines) and the $\epsilon$-fixed sweeping (red dashed lines).
For a given adiabatic parameter $\epsilon$, $B(t)$ changes differently with total particle number $N$ since the energy spectra are different with $N$.
It is shown that, for the same $\epsilon$, as $N$ is getting larger, the required total evolution duration $T$ is longer.

\begin{figure}[!htp]
 \includegraphics[width=0.6\columnwidth]{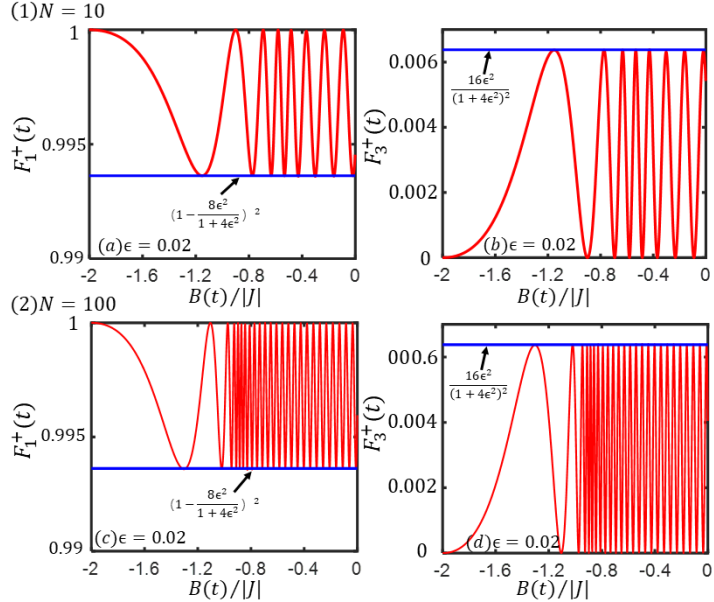}
   \centering\caption{\label{Fig7} (a),(c) The dynamics of fidelity in the instantaneous ground state $F_1$ with $\epsilon = 0.02$ for $N=10$ and $N=100$ respectively.
  (b),(d) The dynamics of fidelity in the instantaneous ground state $F_3$ with $\epsilon = 0.02$ for $N=10$ and $N=100$ respectively.}
\end{figure}

Finally, we demonstrate the time-evolution of the fidelities for $N=10$ and $N=100$.
When $\epsilon$ is small, the population only occupy between the ground state $|{\phi_{E_{1}}^{+}(t)}\rangle$ and the second excited state $|{\phi_{E_{3}}^{+}(t)}\rangle$, which is coincide with our assumption.
The fidelity $F_{1}^{+}(t)$ and the fidelity $F_{3}^{+}(t)$ also satisfy the two inequations, i.e., $ F_{1}^{+}(t)\geq \left(1-\frac{8{\epsilon}^2}{1+4{\epsilon}^2}\right)^{2}$ and $F_{3}^{+}(t) \leq \frac{16{\epsilon}^2}{(1+4{\epsilon}^2)^2}$.
The numerical results confirm our analytical derivation of inequality~(\ref{Eq:a1}), see Fig.~\ref{Fig7}.

\section{Robustness}
In realistic experiments, a systematic bias due to experimental imperfections or a time-dependent random bias caused by stochastic noise may break the parity symmetry of the system.
Here, we investigate the influences of bias on the symmetry-dependent adiabatic evolution.

\begin{figure}[!htp]
 \includegraphics[width=0.6\columnwidth]{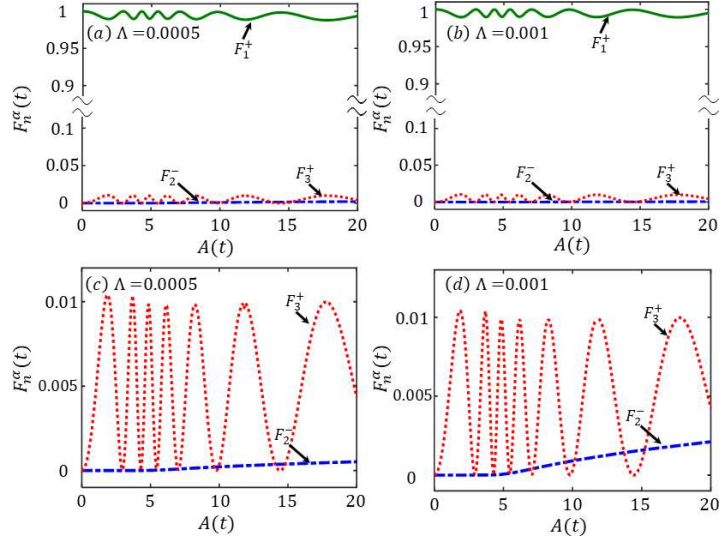}
   \centering\caption{\label{Fig8}(color online). Influences of bias in single-particle system when $\epsilon=0.05$. The evolution of fidelities under (a) $\Lambda=0.0005$ and (b) $\Lambda=0.001$.
  The enlarged details for $F_2^{-}$ and $F_3^{+}$ under (c) $\Lambda=0.0005$ and (d) $\Lambda=0.001$.}
\end{figure}

First, we discuss the systematic bias due to experimental imperfections, which is independent on time.
For the first example of single particle, the bias may cause the imbalance between the two wells, and the system Hamiltonian can be described by
\begin{equation}
\hat{H}_{\textrm{S}}(x,t) = -\frac{{\hbar}^{2}}{2m}\frac{{\partial}^2}{{\partial{x}^2}} + \frac{1}{2}m\omega {x}^{2}+A(t)e^{-\frac{{x}^2}{2d^{2}}} + \Lambda x,
\end{equation}
where $\Lambda$ corresponds to a gradient that imbalance the double-well.
We show the influence of bais in single-particle system in Fig.~\ref{Fig8}.
When $\Lambda$ is small, even though the probability of transferring to the first excited state with odd-parity $|{\phi_{E_2}^{-}}\rangle$ is no longer zero, the transition from $|{\phi_{E_1}^{+}}$ to $|{\phi_{E_3}^{+}}\rangle$ still dominates.

For the second example of transverse field Ising model, the bias corresponds to the longitude field. The system Hamiltonian becomes
\begin{equation}
\hat{H}_{T}' (t) = {B(t)} \sum ^{N}_{i=1}\sigma ^{x}_{i}+J\sum^{N-1}_{i<j}\sigma ^{z}_{i}\sigma ^{z}_{j} + \eta\sum^N_{j=1}\sigma^{z}_{j},
\end{equation}
with $\eta$ the bias.
Similarly, when $\eta$ is small enough, the signature of symmetry-dependent adiabatic evolution still preserve.

\begin{figure}[!htp]
 \includegraphics[width=0.6\columnwidth]{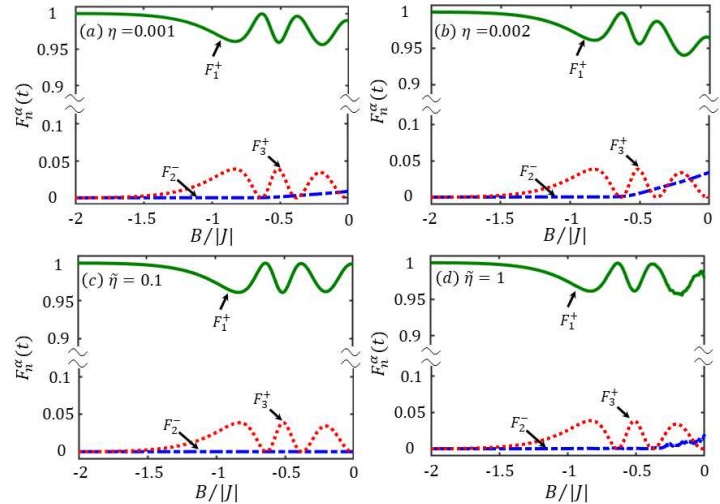}
   \centering\caption{\label{Fig9}(color online). Influences of bias in transverse field Ising system when $\epsilon=0.05$. (a),(b) The evolution of fidelities in the presence of a fixed bias (a) $\eta=0.001$ and (b) $\eta=0.002$. (c),(d) The evolution of fidelities in the presence of a randomly fluctuating bias with maximal amplitude of (c) $\tilde{\eta}=0.1$ and (d) $\tilde{\eta}=1$. Here, $N=10$ and $J=-0.05$.}
\end{figure}

Next, we turn to discuss the influence of time-dependent random fluctuating bias in transverse field Ising model. In this case, the bias fluctuates randomly with time, $\eta\rightarrow\eta(t)$ with $\overline{\eta(t)}=0$ and $\eta(t)\in\left[-\tilde{\eta},\tilde{\eta}\right]$, as shown in Fig.~\ref{Fig9}.
We find that, the symmetry-dependent adiabatic evolution can occur even when the strength of the randomly fluctuating bias is much larger compared with Ising interaction strength $J$.

Our numerical simulation clearly indicates that, even though the bias breaks the parity symmetry, the symmetry-dependent adiabatic evolution can still tolerant small experimental imperfections and random experimental noises, which is robust and feasible in realistic experiments.

\section{Summary and discussions}

We have studied the time-evolution dynamics in a slowly driven system with degeneracy change and time-independent symmetry.
Due to the commutativity between the symmetry and the Hamiltonian, we prove that the population transition is protected by the symmetry.
In further, although the conventional adiabatic condition becomes invalid, we derive a symmetry-dependent adiabatic condition (SDAC).
According to the SDAC, even if there is no energy gap between neighboring eigenstates, we find the existence of adiabatic evolutions.
%
%
By designing proper sweeping processes according to the SDAC, our numerical results confirm the existence of symmetry-protected adiabatic evolutions in both single- and many-particle quantum systems.
Our study will not only deepen the understandings of quantum adiabatic evolution and SSB transitions, but also provide promising applications ranging from quantum state engineering, topological Thouless pumping to quantum computing.

In particular, our findings can be tested via the techniques in quantum annealing.
Our first example can be realized by the quantum annealing of a single superconducting flux qubit~\cite{Johnson2011,Chancellor2017} by switching off the energy bias.
Our second example can be implemented by the quantum annealing in a programmable D-wave system~\cite{Boixo2013} from transverse field limit to Ising interaction limit in the absence of local fields.
Our work is very much related to the recent experiment work for probing quantum criticality and symmetry breaking via Dysprosium atoms, it maybe can provide a platform to verify our work~\cite{Makhalov2019}.
Even if there exist weak symmetry-breaking sources, such as static bias and stochastic noise, the population transfer may be still dominated by the transitions between states with same symmetry.
This means that the symmetry-dependent adiabatic evolution is robust against weak symmetry-breaking sources.

\bigskip
\textbf{Acknowledgments}
\smallskip

M. Zhuang and J. Huang contribute equally to this work. This work is supported by the National Natural Science Foundation of China (NNSFC) under Grants Grants No. 11874434, 11574405 and 11704420. Y. K. was partially supported by International Postdoctoral Exchange Fellowship Program (No. 20180052).

\setcounter{equation}{0}
\renewcommand{\theequation}{A\arabic{equation}}

~\\
\noindent{\bf Appendix A. Derivation of the symmetry-protected transition}
~\\

In this section, we give the proof of Eqs.~(1) and (2), which describe the symmetry-protected transition.
Assume the Hamiltonian $\hat{H}(\textbf{R}(t))=\sum_{i=1}^{K} R_i(t)\hat{H}_i$ with $K$ time-varying parameters
$\textbf{R}(t)=[R_1 (t),R_2 (t),R_3 (t),...,R_K(t)]$,
and has at least one time-independent symmetry $\hat{Y}$ that commutes with the Hamiltonian $\hat{H}$: $[\hat{Y},\hat{\textrm{H}}(\textbf{R}(t))]=0$.
Thus, the operator $\hat{Y}$ and the Hamiltonian $\hat{H}(\textbf{R}(t))$ have a set of simultaneous eigenstates: $\{|{\phi_{E_n}^{\lambda_{\alpha}} (\textbf{R}(t))}\rangle\}$.
Here, $E_{n}$ and $\lambda_{\alpha}$ stand for $n$-th and $\alpha$-th eigenvalues of $\hat{H}(\textbf{R}(t))$ and $\hat{Y}$, respectively.

We write the eigen-equations,
\begin{equation}\label{Eq:Hameigen}
\hat{H}(\textbf{R}(t)) {|{\phi_{E_n}^{\lambda_{\alpha}} (\textbf{R}(t))}\rangle}=E_{n}(\textbf{R}(t)) {|{\phi_{E_n}^{\lambda_{\alpha}} (\textbf{R}(t))}\rangle},
\end{equation}
and
\begin{equation}\label{Eq:Invarienteigen}
\hat{Y} {|{\phi_{E_n}^{\lambda_{\alpha}} (\textbf{R}(t))}\rangle}={\lambda}_{\alpha} {|{\phi_{E_n}^{\lambda_{\alpha}}, (\textbf{R}(t))}\rangle},
\end{equation}
in which
\begin{eqnarray}\label{Eq:H_eigenvalue}
E_{1}(\textbf{R}(t))\leqslant E_{2}(\textbf{R}(t))...\leqslant E_{n}(\textbf{R}(t))...\leqslant E_{N}(\textbf{R}(t)). \nonumber
\end{eqnarray}
%
with $n,\alpha=1,...,N$.
%

By differentiating Eq.~(\ref{Eq:Invarienteigen}) with respect to time, we obtain
\begin{equation}\label{Eq:partial_Invarient1}
\frac{\partial{}}{\partial{t}}\left[\hat{Y} {|{{\phi_{E_n}^{\lambda_{\alpha}} (\textbf{R}(t))}}\rangle}\right]=\hat{Y}\frac{\partial{}}{\partial{t}} {|{{\phi_{E_n}^{\lambda_{\alpha}} (\textbf{R}(t))}}\rangle}= \lambda_{\alpha} {|{{\dot{\phi}_{E_n}^{\lambda_{\alpha}} (\textbf{R}(t))}}\rangle},
\end{equation}
where $|{\dot{\phi}_{E_n}^{\lambda_{\alpha}} (\textbf{R}(t))}\rangle=\frac{\partial}{\partial{t}} |{\phi_{E_n}^{\lambda_{\alpha}} (\textbf{R}(t))}\rangle$.
By taking the inner product with ${\langle{{\phi_{E_m}^{\lambda_{\beta}} (\textbf{R}(t))}}|}$, we obtain
\begin{equation}\label{Eq:partial_Invarient2}
{\langle{{\phi_{E_m}^{\lambda_{\beta}} (\textbf{R}(t))}}|}{\frac{\partial}{\partial{t}}}|{
 {{{\phi}_{E_n}^{\lambda_{\alpha}} (\textbf{R}(t))}}}\rangle(\lambda_{\beta}-\lambda_{\alpha})=0.
\end{equation}
Thus, if ${\lambda_{\beta}}\neq{\lambda_{\alpha}}$, the above equation requests
\begin{equation}\label{Eq:partial_Invarient3}
{\langle{{\phi_{E_m}^{\lambda_{\beta}} (\textbf{R}(t))}}|}{\frac{\partial}{\partial{t}}}|{
 {{{\phi}_{E_n}^{\lambda_{\alpha}} (\textbf{R}(t))}}}\rangle=0.
\end{equation}
On the other hand, the time-evolution of the eigenstate $|{{{\phi}_{E_n}^{\lambda_{\alpha}}
(\textbf{R}(t))}}\rangle$ obeys the Schr\"{o}dinger equation,
\begin{equation}\label{Eq:ShorEigen}
 i\hbar\frac{\partial}{\partial{t}}|{{{\phi}_{E_n}^{\lambda_{\alpha}}
(\textbf{R}(t))}}\rangle=\hat{H}(\textbf{R}(t)) |{{{\phi}_{E_n}^{\lambda_{\alpha}}
(\textbf{R}(t))}}\rangle.
\end{equation}
Thus, substituting Eq.~(\ref{Eq:ShorEigen}) into the Eq.~(\ref{Eq:partial_Invarient3}), we have
\begin{eqnarray}\label{Eq:SPT}
H_{mn}^{\beta\alpha}(t)\! = \!{\langle{\phi_{E_m}^{\lambda_{\beta}}\!(\textbf{R}(t))}|} \hat{H}\!(\textbf{R}(t)) {|{\phi_{E_n}^{\lambda_{\alpha}}\!(\textbf{R}(t))}\rangle} \!=\!0,
\end{eqnarray}
for ${{\lambda}_{\beta}\!\neq \!{\lambda}_{\alpha}}$.
The symmetry protected transition in the main text is proved.

~\\
\noindent{\bf Appendix B. Derivation of the symmetry-dependent adiabatic condition}
~\\

In this section, we give the detailed derivation of the symmetry-dependent adiabatic condition (SDAC) [Eq.~(3) in the main text].
%
%
%
We start from the time-dependent Schr\"{o}dinger equation,
\begin{equation}\label{Eq:Schor_Eq}
i\hbar\frac{\partial{|{\Psi(\textbf{R}(t))}}\rangle}{\partial{t}} =\hat{H}(\textbf{R}(t)) |{\Psi(\textbf{R}(t))}\rangle.
\end{equation}
We assume the system is driven from an instantaneous eigenstate ${|{\phi_{E_m}^{\lambda_{\beta}} (\textbf{R}(t))}\rangle}$ in a degenerate subspace $\mathbb{H}_{\emph{m}}(\textbf{R}(t))$, in which each eigenstate has different symmetry (i.e. $\lambda_{i}\neq\lambda_{j}$ if $i\neq j$ for $i,j=\{1,2,\cdots,d_m\}$).
At any instant, the instantaneous state $|{\Psi(\textbf{R}(t))}\rangle$ can be expanded in terms of the complete basis of ${|{\phi_{E_n}^{\lambda_{\alpha}} (\textbf{R}(t))}\rangle}$,
\begin{equation}\label{Eq:Evolved_state1}
|{\Psi(\textbf{R}(t))} \rangle=\sum_{n\alpha} a_{n}^{\alpha}(t) \textrm{exp}\left[\frac{1}{i\hbar}\int_{0}^{t} E_{n}(\textbf{R}(t'))dt'\right] {|{\phi_{E_n}^{\lambda_{\alpha}} (\textbf{R}(t))}\rangle},
\end{equation}
where $\sum_{n\alpha}$ stands for the sum over all possible combination of $|{\phi_{E_n}^{\lambda_{\alpha}}\rangle}$.
Inserting the above expansion~(\ref{Eq:Evolved_state1}) into the Schr\"{o}dinger equation~(\ref{Eq:Schor_Eq}), we obtain
\begin{eqnarray}\label{Eq:Schor_Eq1}
\sum_{n\alpha} \textrm{exp}\left[\frac{1}{i\hbar}\int_{0}^{t} E_{n}(\textbf{R}(t'))dt'\right]\left\{\dot {a}_{n}^{\alpha}(t)+{a}_{n}^{\alpha}(t)\frac{\partial}{\partial{t}}\right\}
{|{\phi_{E_n}^{\lambda_{\alpha}} (\textbf{R}(t))}\rangle}=0.
\end{eqnarray}
%
Taking the inner product with ${\langle{\phi_{E_m}^{\lambda_{\beta}}(\textbf{R}(t))}|} \textrm{exp}\left[-\frac{1}{i\hbar}\int_{0}^{t} E_{m}\left(\textbf{R}(t')\right)dt'\right]$,
%
%
the differential equation for the coefficients are
%
%
\begin{eqnarray}\label{Eq:Coefficients1}
\dot{a}_{m}^{\beta}(t)=-\sum_{n\alpha} a_{n}^{\alpha}(t) \textrm{exp}\left\{\frac{i}{\hbar}\int_{0}^{t} \Delta_{nm}(t')dt'\right\} {\langle{\phi_{E_m}^{\lambda_{\beta}} (\textbf{R}(t))}|}{{\dot{\phi}_{E_n}^{\lambda_{\alpha}} (\textbf{R}(t))}\rangle}. \nonumber\\
\end{eqnarray}
For briefness, we have introduced $\Delta_{nm}(t)=E_{n}(\textbf{R}(t))-E_{m}(\textbf{R}(t))$. By using the result of Eq.(2) in the main text, substituting Eq.~(\ref{Eq:partial_Invarient3}) into Eq.~(\ref{Eq:Coefficients1}), we have
%
%
%
\begin{eqnarray}\label{Eq:Coefficients}
\dot{a}_{m}^{\beta}(t)
&=&\!\!-\sum_{n\alpha} \!\! a_{n}^{\alpha}(t)\textrm{exp}\!\left\{\frac{i}{\hbar}\int_{0}^{t} \Delta_{nm}(t')dt'\right\}
{\langle{\phi_{E_m}^{\lambda_{\beta}} (\textbf{R}(t))}|}
{{\dot{\phi}_{E_n}^{\lambda_{\alpha}} (\textbf{R}(t))}\rangle} {\delta_{{\lambda_{\alpha}},{\lambda_{\beta}}}}\nonumber\\
\!&=&\!-\sum_{n}  a_{n}^{\beta}(t)\textrm{exp} \left\{\frac{i}{\hbar}\int_{0}^{t} \Delta_{nm}(t')dt'\right\}
{\langle{\phi_{E_m}^{\lambda_{\beta}} (\textbf{R}(t))}|}
{{\dot{\phi}_{E_n}^{\lambda_{\beta}} (\textbf{R}(t))}\rangle}. \nonumber\\
\end{eqnarray}
Therefore, the Hilbert space of the quantum system can be partitioned into different subspaces according to the eigenvalues of the symmetry operator $\hat{Y}$,
and the transition between the states with different eigenvalues of $\hat{Y}$ is forbidden in the dynamical evolution process.

Without loss of generality, we first derive the SDAC for the degenerate quantum system and then relax it to the non-degenerate cases.
%
According to Eq.~(\ref{Eq:Coefficients}), the transitions between states in the same degenerate subspace $\mathbb{H}_{\emph{n}}(\textbf{R}(t))$ are forbidden if all degenerate energy eigenstate possess different values of ${\lambda_{\beta}}$.
Thus, starting from the initial state $|{\Psi(\textbf{R}(0))}\rangle=|{\phi_{E_{m}}^{\lambda_{\beta}} (\textbf{R}(0))}\rangle$,
the evolved state $|{\Psi(\textbf{R}(t))}\rangle$ is given as
\begin{equation}\label{Eq:ES2}
|{\Psi(\textbf{R}(t))}\rangle=\sum_{n} a_{n}^{\beta}(t) \textrm{exp}\left[{\frac{1}{i\hbar} \int_{0}^{t}E_{n}(\textbf{R}(t'))dt'}\right] {|{\phi_{E_{n}}^{\lambda_{\beta}}(\textbf{R}(t))}\rangle}.
\end{equation}
From the time-dependent Schr\"{o}dinger equation, the differential equation for the coefficients in Eq.~(\ref{Eq:ES2}) can be written as
%
%
\begin{eqnarray}\label{Eq:Coefficients3}
&\dot{a}_{m}^{\beta}(t)&=-{a}_{m}^{{\beta}}(t) {\langle{\phi_{E_{m}}^{\lambda_{\beta}}(\textbf{R}(t))}|} {{\dot{\phi}_{E_{m}}^{\lambda_{\beta}}(\textbf{R}(t))}\rangle}\nonumber \\
&-&\!\!\!\!\sum_{n\neq m }{a}_{n}^{{\beta}}(t) \textrm{exp}\left\{\frac{i}{\hbar}\int_{0}^{t} \Delta_{nm}(t')dt'\right\}
{\langle{\phi_{E_{m}}^{\lambda_{\beta}}(\textbf{R}(t))}|}
{{\dot{\phi}_{E_{n}}^{\lambda_{\beta}}(\textbf{R}(t))}\rangle}. \nonumber \\
\end{eqnarray}
Hereafter, we choose properly such that ${\langle{\phi_{E_{m}}^{\lambda_{\alpha}}(\textbf{R}(t))}|} {{\dot{\phi}_{E_{m}}^{\lambda_{\alpha}}(\textbf{R}(t))}\rangle}=0$~\cite{DMTong2007}.
Thus, to ensure the evolved state always stays in the instantaneous eigenstate ${|{\phi_{E_{m}}^{\lambda_{\beta}}(\textbf{R}(t))}\rangle}$, the following adiabatic condition should be satisfied,
\begin{eqnarray}\label{Eq:SAC}
\epsilon(t)=\max_{\{n\}}\left\{ \left| \frac{\hbar{\langle{\phi_{E_{m}}^{\lambda_{\beta}} (\textbf{R}(t))}|
{\dot{\phi}_{E_{n}}^{\lambda_{\beta}} (\textbf{R}(t))}\rangle}}{\Delta_{mn}(t)} \right|\right\} \ll 1~~\textrm{with}~~m\neq n,
\end{eqnarray}
%
%
so that the second term in Eq.~(\ref{Eq:Coefficients3}) can be dropped. Here, $E_m$ and $E_n$ stand for the instantaneous eigenenergies of $\hat{H}(\textbf{R}(t))$, and they respectively belong to different subspaces $\mathbb{H}_{\emph{m}}(\textbf{R}(t))$ and $\mathbb{H}_{\emph{n}}(\textbf{R}(t))$.
Since $i\hbar \frac{\partial}{\partial{t}} =\hat{H}(\textbf{R}(t))$, $H_{mn}^{\alpha\alpha} (t) = i\hbar {\langle{\phi_{E_{m}}^{\lambda_{\alpha}} (\textbf{R}(t))}|
{\dot{\phi}_{E_{n}}^{\lambda_{\alpha}} (\textbf{R}(t))}\rangle}$, and therefore, the SDAC can be written as
%
%
\begin{eqnarray}\label{Eq:AC1}
\epsilon(t)=\max_{\{n\}}\left\{ \left| \frac{H_{mn}^{\beta\beta}(t)}{\Delta_{mn}(t)} \right|\right\} \ll 1~~\textrm{with}~~m\neq n.
\end{eqnarray}
By taking the time derivative on both sides, we also obtain
\begin{eqnarray}\label{Eq:Hameigen_partial}
\frac{ \partial{\hat{H}(\textbf{R}(t))}}{\partial{t}} {|{{\phi}_{E_{n}}^{\lambda_{\beta}} (\textbf{R}(t))}\rangle}+ {\hat{H}(\textbf{R}(t))} {|{\dot{\phi}_{E_{n}}^{\lambda_{\beta}} (\textbf{R}(t))}\rangle} \nonumber \\
=\frac {{\partial{E}_{n}}(\textbf{R}(t))}{\partial{t}} {|{\phi_{E_{n}}^{\lambda_{\beta}} (\textbf{R}(t))}\rangle}+ {{E_{n}}(\textbf{R}(t))} {|{\dot{\phi}_{E_{n}}^{\lambda_{\beta}} (\textbf{R}(t))}\rangle}.
\end{eqnarray}
Multiplying this equation from the left by ${\langle{\phi_{E_{m}}^{\lambda_{\beta}}(\textbf{R}(t))}|}$, we have
%
%
\begin{equation}\label{Eq:Hameigen_partia2}
\langle{\phi_{E_{m}}^{\lambda_{\beta}}(\textbf{R}(t))}| {\dot{\phi}_{E_{n}}^{\lambda_{\beta}}(\textbf{R}(t))}\rangle
=\frac{\langle{\phi_{E_{m}}^{\lambda_{\beta}}(\textbf{R}(t))}| {\frac{\partial{\hat{{H}}(\textbf{R}(t))}}{\partial {t}}}
 {|{\phi_{E_{n}}^{\lambda_{\beta}}(\textbf{R}(t))}\rangle}} {\Delta_{nm}(t)}.
\end{equation}
Therefore, the SDAC~(\ref{Eq:AC1}) can also be rewritten as
%
\begin{eqnarray}\label{Eq:AC_E1}
\epsilon(t)=\max_{\{n\}}\left\{ \left|{\frac{\hbar\langle{\phi_{E_{m}}^{\lambda_{\beta}}(\textbf{R}(t))}|
{\frac{\partial{\hat{{H}}(\textbf{R}(t))}}{\partial {t}}}|
{{\phi_{E_{n}}^{\lambda_{\beta}}(\textbf{R}(t))}}\rangle}
{[{\Delta_{nm}(t)}]^{2}}}\right|\right\} \ll 1 \quad \textrm{with} \quad m\neq n. \nonumber\\
\end{eqnarray}
~\\
\noindent{\bf Appendix C. Analytic analysis of the symmetry-dependent adiabatic evolution}
~\\

In the following, we show how to perform the adiabatic evolution with gap-dependent sweeping designed according to the SDAC~(\ref{Eq:SAC}).
We start from an initial state $|{{\phi_{E_{1}}^{\lambda_{\beta}}(R(0))}}\rangle$ (which is the instantaneous ground state in non-degenerate regime), and vary one of the system parameter $R(t)$ with time to across the quantum phase transition.
The time-varying parameter $R(t)=R(0)+\int_{0}^{t} \upsilon(t') dt'$, where $\upsilon(t)=\dot{R}(t)$ is the sweeping rate of the parameter.
For the linear sweeping scheme, $\dot{R}(t)=\upsilon$ is a constant.

To achieve fast and efficient adiabatic evolution for the ground state, we propose to vary the parameter $R(t)$ non-linearly to keep the adiabatic parameter $\epsilon$ in Eq.~(\ref{Eq:SAC}) a constant.
Substituting ${\frac{\partial{\hat{{H}}(R(t))}}{\partial {t}}}=\frac{\partial{\hat{{H}}(R(t))}}{\partial {R(t)}}\dot{R}(t)=\frac{\partial{\hat{{H}}(R(t))}}{\partial {R(t)}}\upsilon(t)$ into Eq.~(\ref{Eq:AC_E1}), we can obtain
%
%
\begin{eqnarray}\label{Eq:AC_E2}
\upsilon(t)=\frac{\epsilon}{\hbar} \min_{\{n\}}\!\left\{\!\frac{{\left[\Delta_{1n}(t)\right]^{2}}}{\left|\langle{\phi_{E_{1}}^{\lambda_{\beta}}(R(t))}|\!
{\frac{\partial{\hat{{H}}(R(t))}}{\partial {R(t)}}}\!|
{{\phi_{E_{n}}^{\lambda_{\beta}}(R(t))}}\rangle\right|}\!\right\}\!.
\end{eqnarray}
Since we start from the ground state, $a_{1}^{\beta}(0)=1$, and according to Eq.~(\ref{Eq:Coefficients3}), we have
\begin{eqnarray}\label{Eq:Coefficients4}
{a}_{1}^{\beta}(t)=1\!+\!i\hbar\sum_{k > 1 }\!{a}_{k}^{{\beta}}(t) \textrm{exp}\left\{\frac{i}{\hbar}\!\!\int_{0}^{t}\Delta_{k1}(t')dt'\right\}
\frac{{\langle{\phi_{E_{1}}^{\lambda_{\beta}}(R(t))}|}
{{\dot{\phi}_{E_{k}}^{\lambda_{\beta}}(R(t))}\rangle}}{\Delta_{kl}(t')}\nonumber \\
-\!i\hbar\!\sum_{k > 1 }\!\int_{0}^{t}\!\!\textrm{exp}\left\{\frac{i}{\hbar}\!\!\int_{0}^{t'}\Delta_{k1}(t'')dt''\right\}
\frac{{\langle{\phi_{E_{1}}^{\lambda_{\beta}}(R(t'))}|}
{{\dot{\phi}_{E_{k}}^{\lambda_{\beta}}(R(t'))}\rangle}}{\Delta_{kl}(t')}\dot{a}_{k}^{{\beta}}(t')dt'\nonumber \\
-\!i\hbar\!\sum_{k > 1}\!\!\int_{0}^{t}\!\!\textrm{exp}\left\{\frac{i}{\hbar}\int_{0}^{t'}\!\!\Delta_{k1}(t'')dt''\right\}
\frac{\partial}{\partial t'}\!\!\left[\frac{{\langle{\phi_{E_{1}}^{\lambda_{\beta}}(R(t'))}|}
{{\dot{\phi}_{E_{k}}^{\lambda_{\beta}}(R(t'))}\rangle}}{\Delta_{kl}(t')}\right]\!{{a}}_{k}^{{\beta}}(t')dt'\nonumber \\
\end{eqnarray}
In general, the nearest eigenstate of the same symmetry $|{{\phi_{E_{l}}^{\lambda_{\beta}}(R(t))}}\rangle$ determines the sweeping rate~(\ref{Eq:AC_E2}).
If $\epsilon$ is very small, we can make an assumption that the whole process only involve $|{{\phi_{E_{1}}^{\lambda_{\beta}}(R(t))}}\rangle$ and $|{{\phi_{E_{l}}^{\lambda_{\beta}}(R(t))}}\rangle$.
Under this approximation, Eq.~(\ref{Eq:Coefficients4}) can be simplified as
\begin{eqnarray}\label{Eq:Coefficients5}
{a}_{1}^{\beta}(t)=1\!+ i\hbar {a}_{l}^{{\beta}}(t) \textrm{exp}\left\{\frac{i}{\hbar}\int_{0}^{t}\Delta_{l1}(t')dt'\right\}
\frac{{\langle{\phi_{E_{1}}^{\lambda_{\beta}}(R(t))}|}
{{\dot{\phi}_{E_{l}}^{\lambda_{\beta}}(R(t))}\rangle}}{\Delta_{l1}(t')}\nonumber \\
-\!i\hbar  \int_{0}^{t}\!\textrm{exp}\left\{\frac{i}{\hbar}\!\!\int_{0}^{t'}\!\!\Delta_{l1}(t'')dt''\right\}
\frac{{\langle{\phi_{E_{1}}^{\lambda_{\beta}}(R(t'))}|}
{{\dot{\phi}_{E_{l}}^{\lambda_{\beta}}(R(t'))}\rangle}}{\Delta_{l1}(t')}\dot{a}_{l}^{{\beta}}(t')dt'\nonumber \\
-\!i\hbar \int_{0}^{t} \textrm{exp}\left\{\frac{i}{\hbar}\!\!\int_{0}^{t'}\! \!\Delta_{l1}(t'')dt''\right\}
\frac{\partial}{\partial t'}\left[\frac{{\langle{\phi_{E_{1}}^{\lambda_{\beta}}(R(t'))}|}
{{\dot{\phi}_{E_{l}}^{\lambda_{\beta}}(R(t'))}\rangle}}{\Delta_{l1}(t')}\right]\!{{a}}_{l}^{{\beta}}(t')dt'\nonumber \\
\end{eqnarray}
Noting that 
$\left|\textrm{exp}\left\{\frac{i}{\hbar}\int_{0}^{t} \!\Delta_{l1}(t')dt'\right\}\right|=1$,
and the adiabatic parameter $\epsilon(t)=\hbar\left| \frac{{\langle{\phi_{E_{1}}^{\lambda_{\beta}}(R(t'))}|}{{\dot{\phi}_{E_{l}}^{\lambda_{\beta}}(R(t'))}\rangle}}{\Delta_{l1}(t')}\right|=\epsilon$ is time-independent, so that $\frac{\partial}{\partial t'}\left[\frac{{\langle{\phi_{E_{1}}^{\lambda_{\beta}}(R(t'))}|}
{{\dot{\phi}_{E_{l}}^{\lambda_{\beta}}(R(t'))}\rangle}}{\Delta_{l1}(t')}\right]=0$.
Thus, we have
\begin{eqnarray}\label{Eq:Nonequal1b}
\left|{a}_{l}^{{\beta}}(t) \textrm{exp}\left\{\frac{i}{\hbar}\!\int_{0}^{t} \!\!{\Delta_{l1}(t')}dt'\right\}
\frac{{\langle{\phi_{E_{1}}^{\lambda_{\beta}}(R(t))}|}{{\dot{\phi}_{E_{l}}^{\lambda_{\beta}}(R(t))}\rangle}}{{\Delta_{l1}(t)}}\right|\nonumber\\
=\left|{a}_{l}^{{\beta}}(t) \textrm{exp}\left\{\frac{i}{\hbar}\!\int_{0}^{t} \!\!{\Delta_{l1}(t')}dt'\right\} \frac{\epsilon}{\hbar} \right|
\leq
\left|{a}_{l}^{{\beta}}(t)\right|\frac{\epsilon}{\hbar}
\end{eqnarray}
and
%
\begin{eqnarray}\label{Eq:Nonequal2b}
\left|\!\int_{0}^{t} \!\!\textrm{exp}\left\{\frac{i}{\hbar}\!\int_{0}^{t'}\!\!\Delta_{l1}(t'')dt''\right\}\frac{{\langle{\phi_{E_{1}}^{\lambda_{\beta}}(R(t'))}}
{{\dot{\phi}_{E_{l}}^{\lambda_{\beta}}(R(t'))}\rangle}}{\Delta_{l1}(t')}{\dot{a}}_{l}^{{\beta}}(t')dt'\right|
\leq\!\!\frac{\epsilon}{\hbar}\left|\int_{0}^{t}\!\!{\dot{a}}_{l}^{{\beta}}(t')dt'\right|\nonumber\\
\end{eqnarray}
and
\begin{eqnarray}\label{Eq:Nonequal4b}
\int_{0}^{t}\!\!\textrm{exp}\left\{\frac{i}{\hbar}\int_{0}^{t'} {\Delta_{l1}(t'')}dt''\right\}\frac{\partial}{\partial t'}\left[\frac{{\langle{\phi_{E_{1}}^{\lambda_{\beta}}(R(t'))}|}
{{\dot{\phi}_{E_{l}}^{\lambda_{\beta}}(R(t'))}\rangle}}{{\Delta_{l1}(t')}}\right]
{{a}}_{l}^{{\beta}}(t')dt'=0\nonumber\\
\end{eqnarray}
From Eqs.~(\ref{Eq:Nonequal1b})-~(\ref{Eq:Nonequal4b}) and $\left|\int_{0}^{t}\!{\dot{a}}_{l}^{{\beta}}(t')dt'\right|=\left|{{a}}_{l}^{{\beta}}(t)\right|$ ,we can obtain
\begin{eqnarray}\label{Eq:bound}
1-2\epsilon \left|{a}_{l}^{\beta}(t)\right| \le \left|{a}_{1}^{\beta}(t)\right|.
\end{eqnarray}
Due to probability conservation
\begin{eqnarray}\label{Eq:bound2}
 \left|{a}_{1}^{\beta}(t)\right|^{2}+\left|{a}_{l}^{{\beta}}(t)\right|^2 = 1,
\end{eqnarray}
we have
\begin{eqnarray}\label{Eq:bound3}
 \left|{a}_{1}^{\beta}(t)\right|^{2}= 1-\left|{a}_{l}^{{\beta}}(t)\right|^2.
\end{eqnarray}
Squaring Eq.~(\ref{Eq:bound}), and substituting Eq.~(\ref{Eq:bound3}) into it, we get
\begin{eqnarray}\label{Eq:bound4}
\left|{a}_{1}^{\beta}(t)\right|^2 &\geq& 1-4\left|{a}_{l}^{{\beta}}(t)\right| \epsilon+ 4\left|{a}_{l}^{{\beta}}(t)\right|^2 {\epsilon}^2 \nonumber \\
1-\left|{a}_{l}^{\beta}(t)\right|^{2}&\geq& 1-4\left|{a}_{l}^{{\beta}}(t)\right| \epsilon+ 4\left|{a}_{l}^{{\beta}}(t)\right|^2 {\epsilon}^2 \nonumber \\
\left|{a}_{l}^{\beta}(t)\right|^{2} &\leq& 4\left|{a}_{l}^{{\beta}}(t)\right| \epsilon-4\left|{a}_{l}^{{\beta}}(t)\right|^2 {\epsilon}^2 \nonumber \\
\left|{a}_{l}^{\beta}(t)\right|^{2} &\leq& \frac{4\left|{a}_{l}^{{\beta}}(t)\right| \epsilon}{(1+4{\epsilon}^2)}\nonumber \\
\end{eqnarray}
Thus, we obtain the inequality for the coefficient $\left|{a}_{l}^{\beta}(t)\right|$, i.e.,
\begin{eqnarray}\label{Eq:bound5}
\left|{a}_{l}^{\beta}(t)\right| \leq \frac{4\epsilon}{(1+4{\epsilon}^2)},
\end{eqnarray}
or
\begin{eqnarray}\label{Eq:bound6}
\left|{a}_{l}^{\beta}(t)\right|^{2} \leq \frac{16{\epsilon}^2}{(1+4{\epsilon}^2)^2}.
\end{eqnarray}
Further, substituting Eq.~(\ref{Eq:bound5}) into Eq.~(\ref{Eq:bound}), we finally get the inequality for coefficient $\left|{a}_{1}^{\beta}(t)\right|$, i.e.,
\begin{eqnarray}\label{Eq:bound7}
\left|{a}_{1}^{\beta}(t)\right| \geq 1-\frac{8{\epsilon}^2}{(1+4{\epsilon}^2)},
\end{eqnarray}
or
\begin{eqnarray}\label{Eq:bound8}
\left|{a}_{1}^{\beta}(t)\right|^2 \geq \left(1-\frac{8{\epsilon}^2}{(1+4{\epsilon}^2)}\right)^2.
\end{eqnarray}
The inequalities~(\ref{Eq:bound5})-~(\ref{Eq:bound8}) only hold when $\epsilon$ is sufficiently small.
Therefore, we find the minimal fidelity of the ground state
\begin{equation}\label{Eq:Sa1}
F_1^{\beta}(t)\gtrsim\left(1\!-\!\frac{8{\epsilon}^2}{1+4{\epsilon}^2}\right)^2\approx 1-16\epsilon^2+O(\epsilon^4).
\end{equation}

While $\epsilon$ becomes larger, the SDAC~(\ref{Eq:SAC}) gradually breaks, higher excited states (with same symmetry) begin to be populated.
Thus, Eq.~(\ref{Eq:Coefficients5}) will not satisfy and the inequalities no longer hold.
In the following, we will show the comparisons between the analytic and the numerical results.
~\\
~\\
\noindent{\bf Appendix D. Parity operator}
~\\

Here, we give a brief introduction about the parity operator $\hat{P}$.
The parity operator $\hat{P}$ is defined as an operation of space/spin inversion.
The parity operator $\hat{P}$ has the following properties
\begin{equation}\label{Eq:parity1}
\hat{P}^{2}=1 , \quad \hat{P}=P^{\dagger}.
\end{equation}
As it turns out, the parity operator $\hat{P}$ can only ever have two eigenvalues $\xi=\pm1$.
The parity eigenvalue equations are given as
\begin{equation}\label{Eq:parity3}
\hat{P}|{\xi}_{even}\rangle=+1|{\xi}_{even}\rangle,
\end{equation}
and
\begin{equation}\label{Eq:parity4}
\hat{P}|{\xi}_{odd}\rangle=-1|{\xi}_{odd}\rangle.
\end{equation}
This implies that the parity eigenstates will either be the same or be the opposite with their original ones under the space/spin inversion.
If the sign doesn't change, the state $|{\xi}_{even}\rangle$ is symmetric under space inversion (called even).
But, if the sign does change, the state $|{\xi}_{odd}\rangle$ is antisymmetric under space inversion (called odd).
For different quantum systems, the parity operator has different definitions, but they share common properties Eq.(\ref{Eq:parity1}).
If the parity operator $\hat{P}$ commutes with the Hamiltonian $\hat{\textrm{H}}(\textbf{R}(t))$ of the system, we called the system has the parity symmetry.

~\\
\textbf{{References}}
~\\

\end{document}